\documentclass[aps,pre,10pt,twocolumn,longbibliography]{revtex4-1}
\usepackage[ascii]{inputenc}
\usepackage{amsmath,amssymb,amsfonts,amsthm}
\usepackage{graphicx}
\usepackage{bbm}
\usepackage[caption=false]{subfig}
\usepackage[pdftex,bookmarks=false,colorlinks=true,linkcolor=blue,
citecolor=blue,filecolor=black,urlcolor=blue]{hyperref}

\providecommand{\ud}{\mathrm{d}}

\graphicspath{{figures/}
{../analysis/plots/analysis_L2048_wave/}}

\begin{document}

\title{Quantum lattice Boltzmann study of random-mass Dirac fermions in one dimension}

\author{Christian B.~Mendl}
\email{mendl@stanford.edu}
\affiliation{Stanford Institute for Materials and Energy Sciences, SLAC National Accelerator Laboratory and Stanford University, Menlo Park, California 94025, USA}
\author{Silvia Palpacelli}
\email{silvia.palpacelli@hyperlean.eu}
\affiliation{Hyperlean S.r.l, Via Giuseppe Verdi 4, 60122, Ancona, Italy}
\author{Alex Kamenev}
\email{kamenev@physics.umn.edu}
\affiliation{W.~I.~Fine Theoretical Physics Institute and School of Physics and Astronomy, University of Minnesota, Minneapolis, Minnesota 55455, USA}
\author{Sauro Succi}
\email{succi@iac.cnr.it}
\affiliation{Istituto Applicazioni Calcolo, CNR, via dei Taurini 19, 00185, Roma, Italy, and\\
Institute for Applied Computational Science, John Paulson school of Engineering and Applied Sciences, Harvard University, Cambridge, Massachusetts 02138, USA}

\begin{abstract}
We study the time evolution of quenched random-mass Dirac fermions in one dimension by quantum lattice Boltzmann simulations. For nonzero noise strength, the diffusion of an initial wave packet stops after a finite time interval, reminiscent of Anderson localization. However, instead of exponential localization we find algebraically decaying tails in the disorder-averaged density distribution. These qualitatively match $\propto x^{-3/2}$ decay, which has been predicted by analytic calculations based on zero-energy solutions of the Dirac equation.
\end{abstract}

\maketitle

\section{Introduction}

It is a great pleasure, let alone honor, to present this contribution on the occasion of Prof.~Norman H.~March 90th Festschrift. Prof.~March made many distinguished contributions across a broad variety of topics in classical and quantum statistical physics; in the following we present a computational investigation along the latter direction, namely the transport properties of random-mass Dirac fermions in $1+1$ dimensions.

Disorder plays an important role in many physical systems, ranging from topological materials \cite{GrothPRL2009, KobayashiPRL2014, MorimotoFurusakiMudryPRB2015, BagretsAltlandKamenevPRL2016} to transport properties affected by impurities, superconductors \cite{SeoNatPhys2014} and glasses \cite{YunkerPRL2010}. In condensed matter physics, a prominent effect of disorder is exponential Anderson localization of the electronic wavefunction \cite{Anderson1958}, which has been experimentally observed in Bose-Einstein condensates \cite{AspectNature2008}. Nevertheless, around critical points there can be transitions away from the localized phase \cite{BalentsFisher1997, SheltonTsvelikPRB1998, MkhitaryanPRL2011}. In one dimension, similarities between these delocalized phases and classical particle motion in a stationary random potential with a variety of diffusion laws \cite{Sinai1982, Bouchaud1990, ComtetDean1998} have been pointed out, including anomalously slow Sinai diffusion $|x| \propto \log(t)^2$ \cite{BagretsAltlandKamenevPRL2016}.

In this work, we study the time evolution dynamics governed by a prototypical random-mass Dirac equation in one dimension, and investigate the fate of an initial Gaussian wave packet. The general framework is similar to a recent related work \cite{YosprakobSuwanna2016}, except for the numerical quantum lattice Boltzmann approach pursued here, and different versions of the Dirac equation. Specifically, using the Majorana representation and projecting upon chiral eigenstates (and setting $\hbar = 1$), the Dirac equation considered here reads
\begin{equation}
\label{eq:dirac}
\big(i \partial_t + i c \sigma^z \partial_x + c^2 m(x) \sigma^y \big) \psi(x,t) = 0,
\end{equation}
where $\psi(x,t)$ is a two-component spinor, $\sigma^{\alpha}$ are the Pauli matrices, $c$ the speed of light, and $m(x)$ is the spatially dependent mass. We model quenched disorder by taking $m(x)$ as a Gaussian white noise random variable with mean $m_0$ and noise strength $\lambda$:
\begin{equation}
\label{eq:random_mass_distr}
\langle (m(x) - m_0) (m(x') - m_0) \rangle = 2 \lambda \delta(x - x').
\end{equation}
The spinor $\psi = (u, d)^T$ consists of the chiral right-moving ($u$) and left-moving ($d$) states. The stationary version of Eq.~\eqref{eq:dirac} (without the time derivative) has been identified as an effective theory in a tight-binding model of spinless fermions \cite{BalentsFisher1997}.

The dynamics governed by \eqref{eq:dirac} conserves total density and energy. For example, the local density
\begin{equation}
\label{eq:rho}
\rho = |\psi|^2 = |u|^2 + |d|^2
\end{equation}
obeys the conservation law
\begin{equation}
\partial_t \rho(x,t) + \partial_x J_{\rho}(x,t) = 0
\end{equation}
with the density current
\begin{equation}
J_{\rho}(x,t) = c \left( |u|^2 - |d|^2 \right).
\end{equation}

We will see in the numerical simulations that $\psi(x,t)$ converges to a stationary state for $\lambda > 0$; this stationary state can thus be compared to the zero-energy solution studied in \cite{BalentsFisher1997}: $\psi(x) = \psi_{\pm}(x) (\begin{smallmatrix}1 \\ \mp 1\end{smallmatrix})$, with the scalar function $\psi_{\pm}(x)$ satisfying
\begin{equation}
\big( \partial_x \pm c m(x) \big) \psi_{\pm}(x) = 0.
\end{equation}
For ``critical'' zero average mass ($m_0 = 0$), this results in the log-normally distributed wavefunction
\begin{equation}
\label{eq:psi_lognormal}
\psi_{\pm}(x) \propto \mathrm{e}^{\pm \int_0^x c m(x') \ud x'},
\end{equation}
which deviates from exponential localization. By a mapping to Liouville field theory, the disorder-averaged spatial correlations of the wavefunction \eqref{eq:psi_lognormal} can be computed analytically \cite{BalentsFisher1997, SheltonTsvelikPRB1998, SteinerPRB1998}, resulting in an \emph{algebraic} (instead of exponential) decay with exponent $-3/2$:
\begin{equation}
\label{eq:psi_alg_decay}
\left\langle \lvert \psi(x) \rvert^2 \lvert \psi(0) \rvert^2 \right\rangle \propto \lvert x \rvert^{-3/2}.
\end{equation}
Thus, disorder in the random mass distribution does not lead to Anderson localization if the average mass is zero.

\section{Quantum lattice Boltzmann method}

Eq.~\eqref{eq:dirac} lends itself to a lattice Boltzmann discretization for the spinor components $u$ and $d$, as observed in \cite{SucciBenzi1993, PalpacelliSucciPRE2008, FillionGourdeauPRL2013}. The propagation step consists of streaming $u$ and $d$ along the $x$-axis with opposite speeds $\pm c$, while the collision step is performed according to the scattering term $c^2 m(x) \sigma^y \psi$. Integrating \eqref{eq:dirac} along the characteristics of $u$ and $d$, respectively, and approximating the collision integral by the trapezoidal rule, the following relations are obtained:
\begin{equation} \label{eq:QLB}
\begin{split}
\hat{u} - u &= \tilde{m} (d+ \hat{d}) / 2\\
\hat{d} - d &= - \tilde{m} (u+ \hat{u}) / 2,
\end{split}
\end{equation}
where $\hat{u} = u(x + \Delta x, t + \Delta t)$, $\hat{d} = d(x - \Delta x, t + \Delta t)$, $\Delta x = c \Delta t$, and $\tilde{m} = c^2 m \Delta t$. Algebraically solving the linear system \eqref{eq:QLB} yields the explicit scheme
\begin{equation} \label{eq:QLB_scheme}
\begin{pmatrix} \hat{u} \\ \hat{d} \end{pmatrix} = \begin{pmatrix} a & b \\ -b & a \end{pmatrix} \begin{pmatrix} u \\ d \end{pmatrix},
\end{equation}
with
\begin{equation*}
a = (1 - \tilde{m}^2/4)/(1 + \tilde{m}^2/4), \quad b = \tilde{m}/(1 + \tilde{m}^2/4).
\end{equation*}
Note that, since $|a|^2 + |b|^2 = 1$, the collision matrix is unitary, thus the method is unconditionally stable and norm-preserving.

\section{Numerical simulation results}

We start from a ``wave packet'' initial state given by
\begin{equation}
\psi(x,0)\equiv \begin{pmatrix} u \\ d \end{pmatrix} = \big(\sqrt{8 \pi} \sigma\big)^{-1/2} \mathrm{e}^{-x^2/4 \sigma^2 } \begin{pmatrix} 1\\ 1 \end{pmatrix},
\end{equation}
with the standard deviation $\sigma$ measuring the width of the wave packet, and the normalization chosen such that $\int_{-\infty}^{\infty} \rho(x,t) \, \ud x = 1$ at $t = 0$. Due to density conservation, this relation holds for all $t$.

Table~\ref{tab:sim_params} lists the simulation parameters in detail. The speed of light $c = \Delta x / \Delta t = 1$, and the physical simulation domain is the interval $[-64,64]$.

\begin{table}[!ht]
\begin{tabular}{clp{0.8\columnwidth}}
$L$        & 2048   & system size (number of grid points) with periodic boundary conditions \\
$\Delta x$ & $1/16$ & grid spacing \\
$\Delta t$ & $1/16$ & time step \\
$\sigma$   & $1$    & standard deviation of initial spinor \\
$n_{\text{runs}}$   & $10^5$ & number of random mass realizations (simulation runs) to compute averages $\langle \dots \rangle$ \\
$n_{\text{cut}}$   & $256$ & cut-off Fourier mode of random mass distribution
\end{tabular}
\caption{Simulation parameters}
\label{tab:sim_params}
\end{table}

Eq.~\eqref{eq:random_mass_distr} suggests to draw a random $m(x_i)$ independently at each grid point $x_i$. However, this would render the simulation sensitive to the grid spacing $\Delta x$. Instead, we draw independent Fourier coefficients up to some cut-off Fourier mode $n_{\text{cut}}$, and then transform to real space to obtain a random mass realization. Thus, the grid resolution is much finer than random mass oscillations. The random mass correlations obtained by this procedure decay on a length scale $x - x' = \Delta x L /(2 n_{\text{cut}})$. This quantity is chosen small compared to the width of the initial wave packet, in order to approximate the delta function in Eq.~\eqref{eq:random_mass_distr}.

\begin{figure}[!ht]
\centering
\includegraphics[width=\columnwidth]{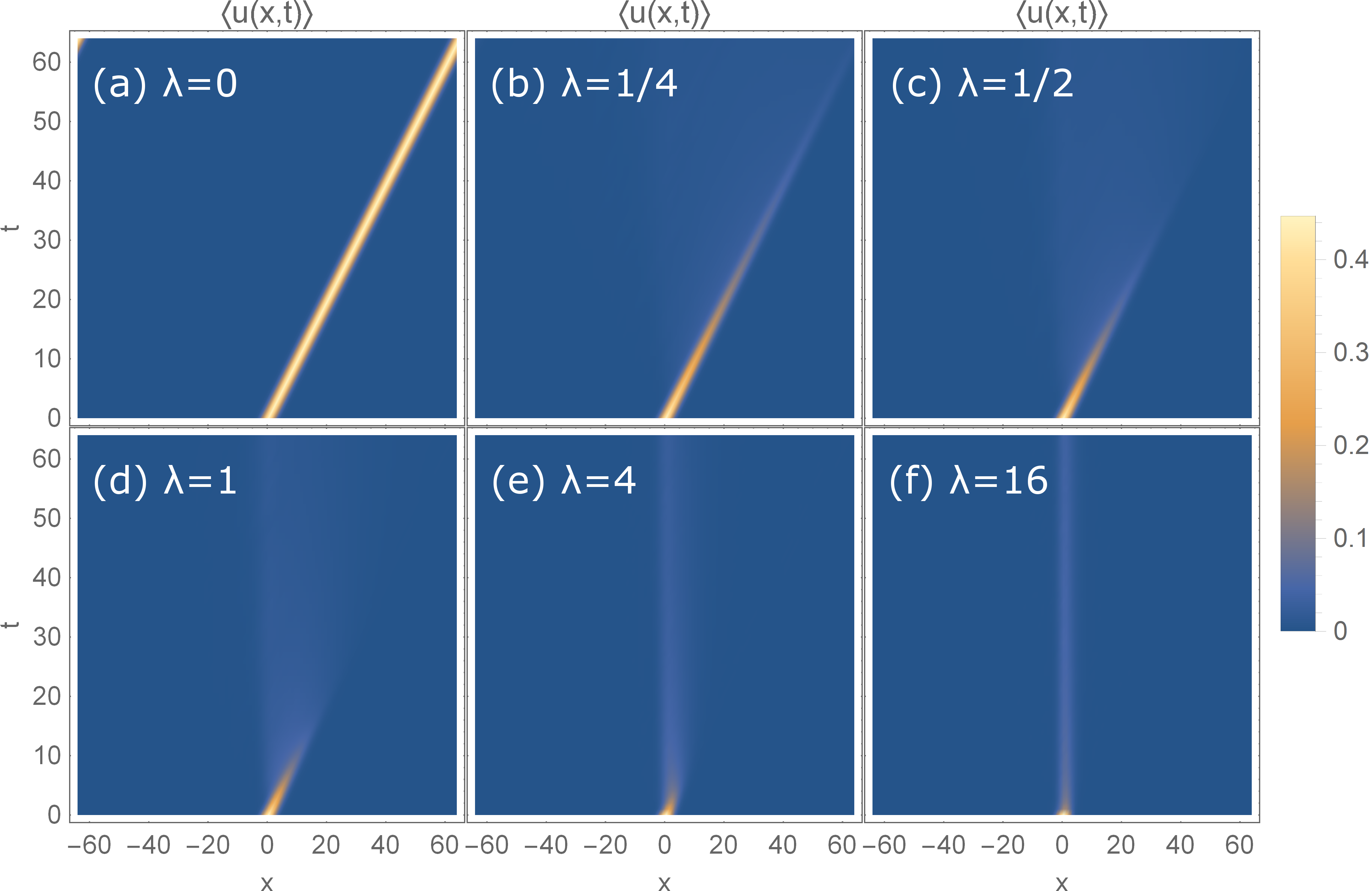}
\caption{Average $\langle u(x,t) \rangle$ profile for increasing noise strength of the random mass distribution, and $m_0 = 0$.}
\label{fig:u_avr}
\end{figure}
Fig.~\ref{fig:u_avr} shows $\langle u(x,t) \rangle$ for various values of $\lambda$, for zero average mass ($m_0 = 0$). In the absence of noise ($\lambda = 0$), there is no scattering term in the Dirac equation, and the $u$ and $d$ waves freely propagate to the right and left, respectively. For $\lambda > 0$, the right-moving ray is continuously diminished over time due to scattering. As $\lambda$ increases, the wave packet remains more and more tied to the origin.

\begin{figure}[!ht]
\centering
\includegraphics[width=\columnwidth]{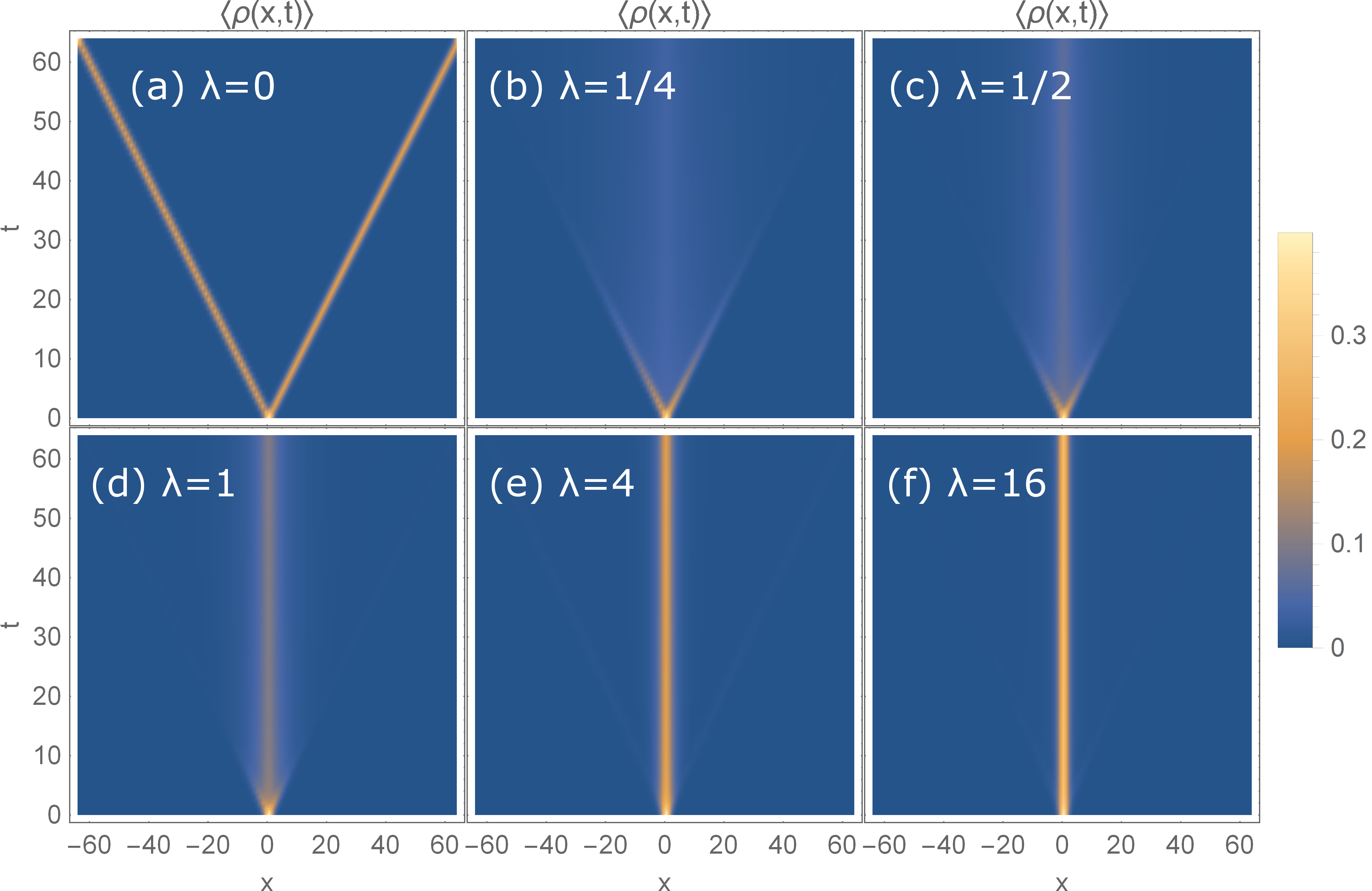}
\caption{Average density $\langle \rho(x,t) \rangle$ for increasing noise strength of the random mass distribution, and $m_0 = 0$.}
\label{fig:rho_avr}
\end{figure}
Fig.~\ref{fig:rho_avr} visualizes the corresponding density profiles $\langle \rho(z,t) \rangle$ for the same simulations. For any $\lambda > 0$, one observes remnant density centered around the origin. The density profile remains stationary at later times.

\begin{figure}[!ht]
\centering
\includegraphics[width=0.85\columnwidth]{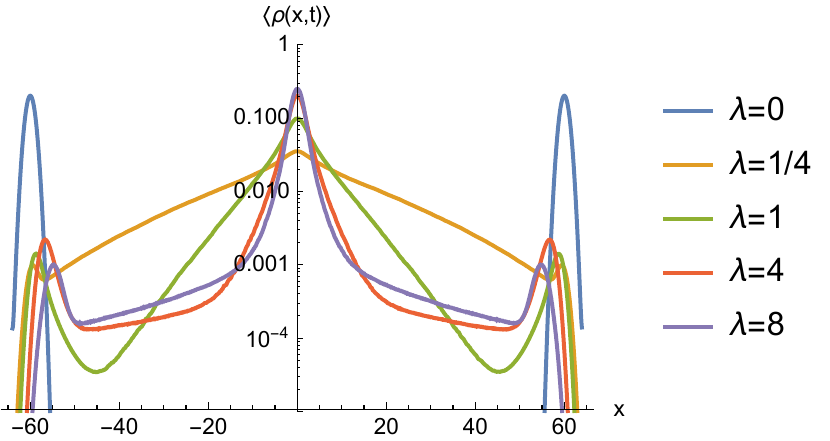}
\caption{Average density $\langle \rho(x,t) \rangle$ at $t = 60$ on a logarithmic scale, for $m_0 = 0$.}
\label{fig:rho_log}
\end{figure}
To analyze the noise-averaged density quantitatively, Fig.~\ref{fig:rho_log} shows the density profile on a logarithmic scale at $t = 60$, when it has (almost) reached stationarity between the left- and right-moving sound peaks around $x \simeq \pm 60$. The density decays exponentially with respect to $|x|$ for $0 < \lambda \lesssim 1$, different from the predicted algebraic decay in Eq.~\eqref{eq:psi_alg_decay}. One explanation could be that the algebraic decay sets in at larger $|x|$. On the other hand, for $\lambda \gtrsim 4$, one observes a transition from exponential to slower-decaying tails. (Note that for the particular initial condition used in our simulations, we find that the density \emph{correlation} between the the origin and $x$ is proportional to the density profile.)

\begin{figure}[!ht]
\centering
\includegraphics[width=0.825\columnwidth]{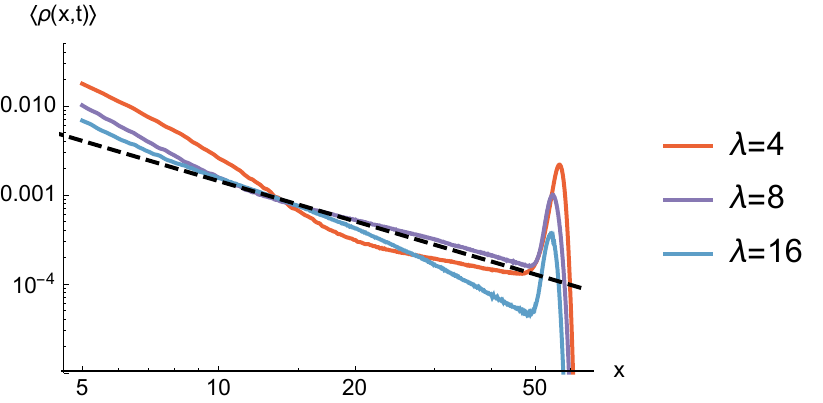}
\caption{Average density $\langle \rho(x,t) \rangle$ at $t = 60$ on a log-log scale, for $m_0 = 0$. For comparison, the black dashed line is $\propto x^{-3/2}$.}
\label{fig:rho_loglog}
\end{figure}
Fig.~\ref{fig:rho_loglog} shows these tails on a log-log scale, which indeed ascertains an algebraic decay at larger $\lvert x \rvert$. Between $20 < x < 45$, the curve for noise strength $\lambda = 4$ decays somewhat slower, the $\lambda = 16$ curve somewhat faster, and the $\lambda = 8$ curve almost exactly as the black dashed $\propto x^{-3/2}$ line based on the theoretical prediction \eqref{eq:psi_alg_decay}.

\begin{figure}[!ht]
\centering
\includegraphics[width=0.625\columnwidth]{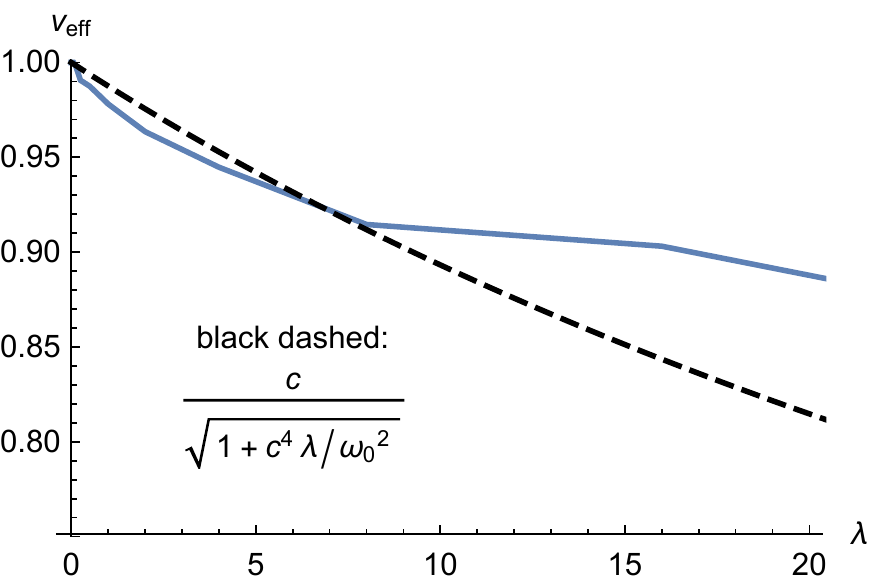}
\caption{Measured sound velocity in dependence of noise strength $\lambda$, for $m_0 = 0$.}
\label{fig:sound_speed}
\end{figure}

The logarithmic scale in Fig.~\ref{fig:rho_log} shows that the outward-moving sound peaks are present also for $\lambda \ge 1$, even though not visible in Fig.~\ref{fig:rho_avr}. The effective sound velocity $v_{\text{eff}}$ (measured via the peak maximum) monotonically decreases with noise strength, as expected (see Fig.~\ref{fig:sound_speed}).

Solutions of the free Dirac equation also solve the Klein-Gordon equation with dispersion relation $\omega^2 = (c k)^2 + \omega_c^2$, where $\omega_c = c^2 m / \hbar$ is the Compton frequency. The corresponding sound speed is therefore
\begin{equation}
v_{\text{KG}} = \partial_k \omega = \frac{c}{\sqrt{1 + (\omega_c / (c k))^2}}.
\end{equation}
The wave number $k$ should be inversely proportional to the spatial extent of the wave packet; thus we approximate $c k \simeq \omega_0$ with $\omega_0 = 2 \pi c/ \sigma$. For the Compton frequency, we use $\sqrt{\lambda}$ as proxy for the mass term, and set $\hbar = 1$ as before. This results in the black dashed curve in Fig.~\ref{fig:sound_speed}, which indeed qualitatively reproduces the measured sound velocity up to $\lambda \lesssim 8$.

\begin{figure}[!ht]
\centering
\includegraphics[width=\columnwidth]{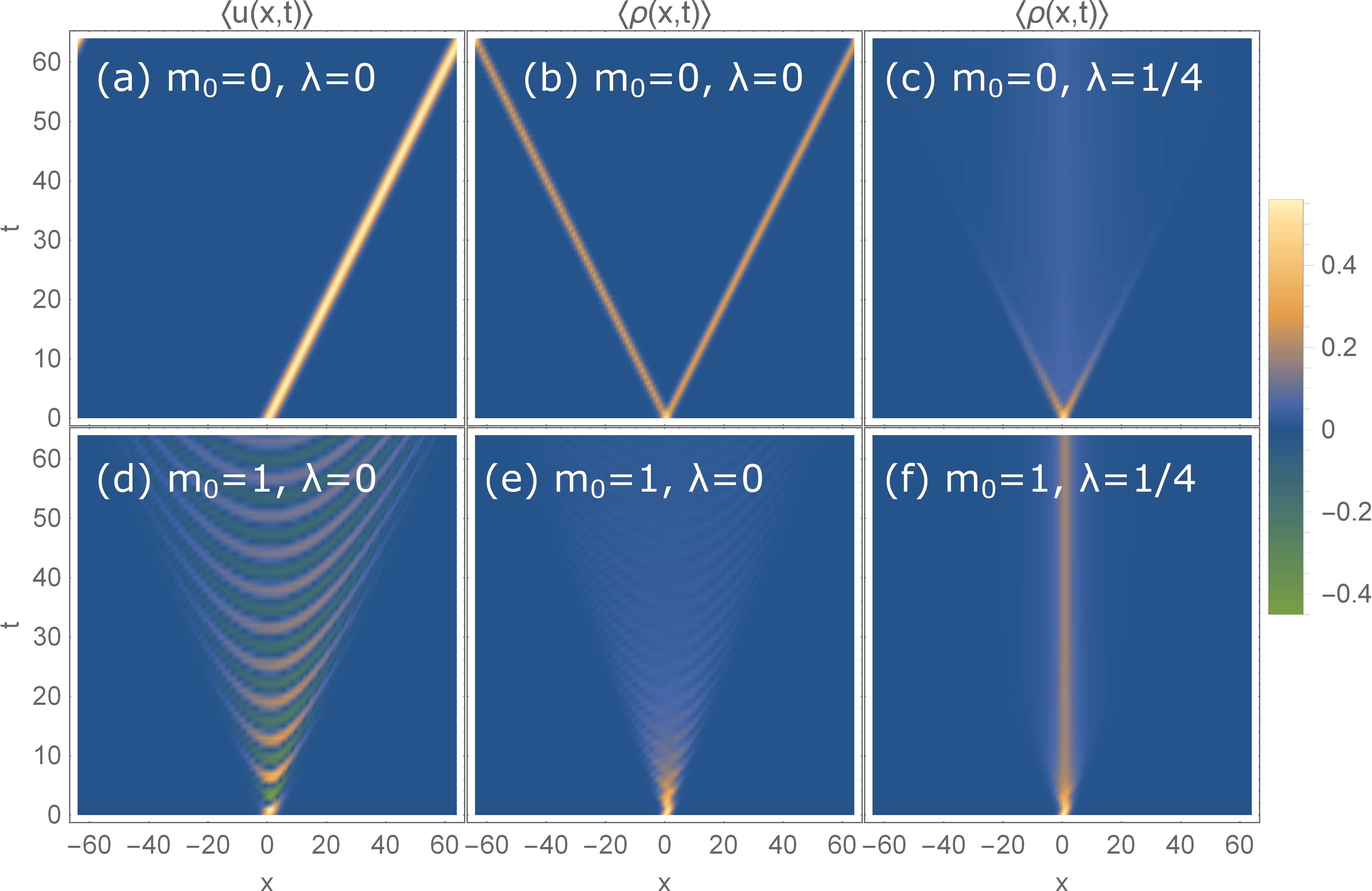}
\caption{Comparison of the $\langle u(x,t) \rangle$ profile and density $\langle \rho(x,t) \rangle$ for $m_0 = 0$ (top row) with $m_0 = 1$ (bottom row).}
\label{fig:u_rho_m0_comparison}
\end{figure}
Tuning away from zero average mass should result in ``conventional'' exponentially localized wavefunctions (see also Eq.~\eqref{eq:psi_lognormal}) at zero-energy. Fig.~\ref{fig:u_rho_m0_comparison} directly compares hitherto $m_0 = 0$ simulations with $m_0 = 1$. Without disorder ($\lambda = 0$), the $u$ (and $d$) component exhibits a parabola-shaped stripe pattern (see Fig.~\ref{fig:u_rho_m0_comparison}d), instead of linear propagation. The corresponding density has a more uniform profile. When including disorder ($\lambda = 1/4$), one notices that the average density is more strongly confined for $m_0 = 1$ (Fig.~\ref{fig:u_rho_m0_comparison}f) than for $m_0 = 0$ (Fig.~\ref{fig:u_rho_m0_comparison}c).

\begin{figure}[!ht]
\centering
\includegraphics[width=0.85\columnwidth]{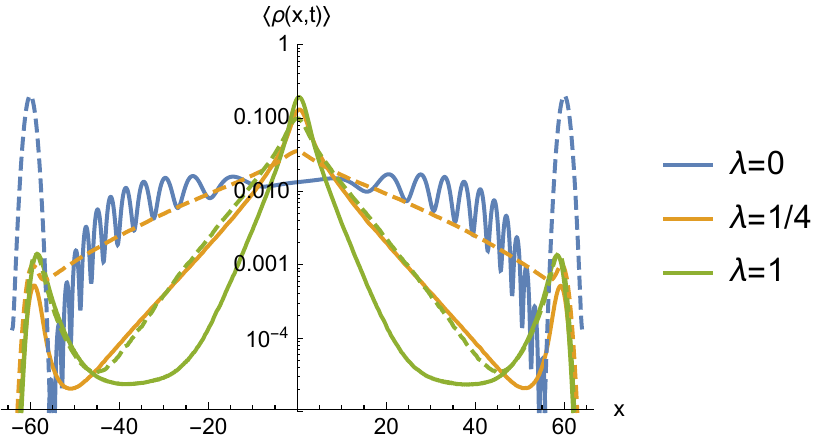}
\caption{Average density $\langle \rho(x,t) \rangle$ for $m_0 = 1$ (solid lines) compared to $m_0 = 0$ (dashed lines, same data as in Fig.~\ref{fig:rho_log}).}
\label{fig:rho_log_m0_comparison}
\end{figure}
This stronger confinement is confirmed in Fig.~\ref{fig:rho_log_m0_comparison}, which compares the densities on a logarithmic scale for $0 \le \lambda \le 1$. Besides the oscillatory pattern at $\lambda = 0$, the density for $m_0 = 1$ decays faster than for $m_0 = 0$ at fixed $\lambda > 0$.

\begin{figure}[!ht]
\centering
\includegraphics[width=0.825\columnwidth]{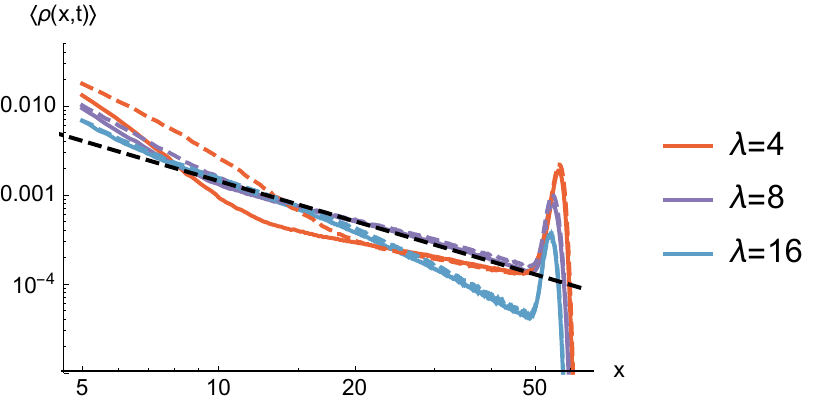}
\caption{Average density $\langle \rho(x,t) \rangle$ on a log-log scale for $m_0 = 1$ (solid lines) compared to $m_0 = 0$ (dashed lines, same data as in Fig.~\ref{fig:rho_loglog}).}
\label{fig:rho_loglog_m0_comparison}
\end{figure}
Fig.~\ref{fig:rho_loglog_m0_comparison} compares the densities on a log-log scale for $\lambda \ge 4$. Somewhat surprisingly, the non-zero average mass $m_0 = 1$ does not affect the algebraic decay, although one would expect exponential decay away from the ``critical'' $m_0 = 0$. An explanation could be that large values of the noise override small changes in the average mass.

\section{Conclusions and outlook}

We have shown that quantum lattice Boltzmann methods can efficiently simulate the real-time dynamics of the single-particle Dirac equation \eqref{eq:dirac} for random-mass fermions in one spatial dimension. Since the quantum lattice Boltzmann scheme is not limited to one-dimensional systems \cite{DellarPRE2011}, for the future it would be interesting to study the transport properties of random-mass fermions in two and three spatial dimensions. Besides analyzing stationary properties, lattice Boltzmann simulations of the Dirac equation could also be used for investigating the time dynamics of out-of-equilibrium systems, including, e.g., thermalization and quasiparticle lifetime, cf.~\cite{SucciEPL2015}. Work along the lines is currently underway.

\begin{acknowledgments}

This work is dedicated to Prof.~Norman H.~March on the occasion of his 90th Festschrift, with our warmest congratulations on an outstanding career and best wishes for more to come in the future.

C.M.\ acknowledges support from the Alexander von Humboldt foundation via a Feodor Lynen fellowship, as well as support from the US Department of Energy, Office of Basic Energy Sciences, Division of Materials Sciences and Engineering, under Contract No.~DE-AC02-76SF00515. A.K.\ was supported by NSF grant DMR-1608238. S.S.\ was supported by the European Research Council under the European Union's Seventh Framework Programme (FP/2007-2013)/ERC Grant Agreement No.~306357 (ERC Starting Grant ``NANO-JETS'').
\end{acknowledgments}


\begin{thebibliography}{21}%
\makeatletter
\providecommand \@ifxundefined [1]{%
 \@ifx{#1\undefined}
}%
\providecommand \@ifnum [1]{%
 \ifnum #1\expandafter \@firstoftwo
 \else \expandafter \@secondoftwo
 \fi
}%
\providecommand \@ifx [1]{%
 \ifx #1\expandafter \@firstoftwo
 \else \expandafter \@secondoftwo
 \fi
}%
\providecommand \natexlab [1]{#1}%
\providecommand \enquote  [1]{``#1''}%
\providecommand \bibnamefont  [1]{#1}%
\providecommand \bibfnamefont [1]{#1}%
\providecommand \citenamefont [1]{#1}%
\providecommand \href@noop [0]{\@secondoftwo}%
\providecommand \href [0]{\begingroup \@sanitize@url \@href}%
\providecommand \@href[1]{\@@startlink{#1}\@@href}%
\providecommand \@@href[1]{\endgroup#1\@@endlink}%
\providecommand \@sanitize@url [0]{\catcode `\\12\catcode `\$12\catcode
  `\&12\catcode `\#12\catcode `\^12\catcode `\_12\catcode `\%12\relax}%
\providecommand \@@startlink[1]{}%
\providecommand \@@endlink[0]{}%
\providecommand \url  [0]{\begingroup\@sanitize@url \@url }%
\providecommand \@url [1]{\endgroup\@href {#1}{\urlprefix }}%
\providecommand \urlprefix  [0]{URL }%
\providecommand \Eprint [0]{\href }%
\providecommand \doibase [0]{http://dx.doi.org/}%
\providecommand \selectlanguage [0]{\@gobble}%
\providecommand \bibinfo  [0]{\@secondoftwo}%
\providecommand \bibfield  [0]{\@secondoftwo}%
\providecommand \translation [1]{[#1]}%
\providecommand \BibitemOpen [0]{}%
\providecommand \bibitemStop [0]{}%
\providecommand \bibitemNoStop [0]{.\EOS\space}%
\providecommand \EOS [0]{\spacefactor3000\relax}%
\providecommand \BibitemShut  [1]{\csname bibitem#1\endcsname}%
\let\auto@bib@innerbib\@empty
\bibitem [{\citenamefont {Groth}\ \emph {et~al.}(2009)\citenamefont {Groth},
  \citenamefont {Wimmer}, \citenamefont {Akhmerov}, \citenamefont
  {Tworzyd\l{}o},\ and\ \citenamefont {Beenakker}}]{GrothPRL2009}%
  \BibitemOpen
  \bibfield  {author} {\bibinfo {author} {\bibfnamefont {C.~W.}\ \bibnamefont
  {Groth}}, \bibinfo {author} {\bibfnamefont {M.}~\bibnamefont {Wimmer}},
  \bibinfo {author} {\bibfnamefont {A.~R.}\ \bibnamefont {Akhmerov}}, \bibinfo
  {author} {\bibfnamefont {J.}~\bibnamefont {Tworzyd\l{}o}}, \ and\ \bibinfo
  {author} {\bibfnamefont {C.~W.~J.}\ \bibnamefont {Beenakker}},\ }\bibfield
  {title} {\enquote {\bibinfo {title} {{Theory of the topological Anderson
  insulator}},}\ }\href {\doibase 10.1103/PhysRevLett.103.196805} {\bibfield
  {journal} {\bibinfo  {journal} {Phys. Rev. Lett.}\ }\textbf {\bibinfo
  {volume} {103}},\ \bibinfo {pages} {196805} (\bibinfo {year}
  {2009})}\BibitemShut {NoStop}%
\bibitem [{\citenamefont {Kobayashi}\ \emph {et~al.}(2014)\citenamefont
  {Kobayashi}, \citenamefont {Ohtsuki}, \citenamefont {Imura},\ and\
  \citenamefont {Herbut}}]{KobayashiPRL2014}%
  \BibitemOpen
  \bibfield  {author} {\bibinfo {author} {\bibfnamefont {K.}~\bibnamefont
  {Kobayashi}}, \bibinfo {author} {\bibfnamefont {T.}~\bibnamefont {Ohtsuki}},
  \bibinfo {author} {\bibfnamefont {K.-I.}\ \bibnamefont {Imura}}, \ and\
  \bibinfo {author} {\bibfnamefont {I.~F.}\ \bibnamefont {Herbut}},\ }\bibfield
   {title} {\enquote {\bibinfo {title} {{Density of states scaling at the
  semimetal to metal transition in three dimensional topological
  insulators}},}\ }\href {\doibase 10.1103/PhysRevLett.112.016402} {\bibfield
  {journal} {\bibinfo  {journal} {Phys. Rev. Lett.}\ }\textbf {\bibinfo
  {volume} {112}},\ \bibinfo {pages} {016402} (\bibinfo {year}
  {2014})}\BibitemShut {NoStop}%
\bibitem [{\citenamefont {Morimoto}\ \emph {et~al.}(2015)\citenamefont
  {Morimoto}, \citenamefont {Furusaki},\ and\ \citenamefont
  {Mudry}}]{MorimotoFurusakiMudryPRB2015}%
  \BibitemOpen
  \bibfield  {author} {\bibinfo {author} {\bibfnamefont {T.}~\bibnamefont
  {Morimoto}}, \bibinfo {author} {\bibfnamefont {A.}~\bibnamefont {Furusaki}},
  \ and\ \bibinfo {author} {\bibfnamefont {C.}~\bibnamefont {Mudry}},\
  }\bibfield  {title} {\enquote {\bibinfo {title} {{Anderson localization and
  the topology of classifying spaces}},}\ }\href {\doibase
  10.1103/PhysRevB.91.235111} {\bibfield  {journal} {\bibinfo  {journal} {Phys.
  Rev. B}\ }\textbf {\bibinfo {volume} {91}},\ \bibinfo {pages} {235111}
  (\bibinfo {year} {2015})}\BibitemShut {NoStop}%
\bibitem [{\citenamefont {Bagrets}\ \emph {et~al.}(2016)\citenamefont
  {Bagrets}, \citenamefont {Altland},\ and\ \citenamefont
  {Kamenev}}]{BagretsAltlandKamenevPRL2016}%
  \BibitemOpen
  \bibfield  {author} {\bibinfo {author} {\bibfnamefont {D.}~\bibnamefont
  {Bagrets}}, \bibinfo {author} {\bibfnamefont {A.}~\bibnamefont {Altland}}, \
  and\ \bibinfo {author} {\bibfnamefont {A.}~\bibnamefont {Kamenev}},\
  }\bibfield  {title} {\enquote {\bibinfo {title} {{Sinai diffusion at quasi-1D
  topological phase transitions}},}\ }\href {\doibase
  10.1103/PhysRevLett.117.196801} {\bibfield  {journal} {\bibinfo  {journal}
  {Phys. Rev. Lett.}\ }\textbf {\bibinfo {volume} {117}},\ \bibinfo {pages}
  {196801} (\bibinfo {year} {2016})}\BibitemShut {NoStop}%
\bibitem [{\citenamefont {Seo}\ \emph {et~al.}(2014)\citenamefont {Seo},
  \citenamefont {Lu}, \citenamefont {Zhu}, \citenamefont {Urbano},
  \citenamefont {Curro}, \citenamefont {Bauer}, \citenamefont {Sidorov},
  \citenamefont {Pham}, \citenamefont {Park}, \citenamefont {Fisk},\ and\
  \citenamefont {Thompson}}]{SeoNatPhys2014}%
  \BibitemOpen
  \bibfield  {author} {\bibinfo {author} {\bibfnamefont {S.}~\bibnamefont
  {Seo}}, \bibinfo {author} {\bibfnamefont {X.}~\bibnamefont {Lu}}, \bibinfo
  {author} {\bibfnamefont {J-X.}\ \bibnamefont {Zhu}}, \bibinfo {author}
  {\bibfnamefont {R.~R.}\ \bibnamefont {Urbano}}, \bibinfo {author}
  {\bibfnamefont {N.}~\bibnamefont {Curro}}, \bibinfo {author} {\bibfnamefont
  {E.~D.}\ \bibnamefont {Bauer}}, \bibinfo {author} {\bibfnamefont {V.~A.}\
  \bibnamefont {Sidorov}}, \bibinfo {author} {\bibfnamefont {L.~D.}\
  \bibnamefont {Pham}}, \bibinfo {author} {\bibfnamefont {T.}~\bibnamefont
  {Park}}, \bibinfo {author} {\bibfnamefont {Z.}~\bibnamefont {Fisk}}, \ and\
  \bibinfo {author} {\bibfnamefont {J.~D.}\ \bibnamefont {Thompson}},\
  }\bibfield  {title} {\enquote {\bibinfo {title} {{Disorder in quantum
  critical superconductors}},}\ }\href {\doibase 10.1038/nphys2820} {\bibfield
  {journal} {\bibinfo  {journal} {Nat. Phys.}\ }\textbf {\bibinfo {volume}
  {10}},\ \bibinfo {pages} {120--125} (\bibinfo {year} {2014})}\BibitemShut
  {NoStop}%
\bibitem [{\citenamefont {Yunker}\ \emph {et~al.}(2010)\citenamefont {Yunker},
  \citenamefont {Zhang},\ and\ \citenamefont {Yodh}}]{YunkerPRL2010}%
  \BibitemOpen
  \bibfield  {author} {\bibinfo {author} {\bibfnamefont {P.}~\bibnamefont
  {Yunker}}, \bibinfo {author} {\bibfnamefont {Z.}~\bibnamefont {Zhang}}, \
  and\ \bibinfo {author} {\bibfnamefont {A.~G.}\ \bibnamefont {Yodh}},\
  }\bibfield  {title} {\enquote {\bibinfo {title} {{Observation of the
  disorder-induced crystal-to-glass transition}},}\ }\href {\doibase
  10.1103/PhysRevLett.104.015701} {\bibfield  {journal} {\bibinfo  {journal}
  {Phys. Rev. Lett.}\ }\textbf {\bibinfo {volume} {104}},\ \bibinfo {pages}
  {015701} (\bibinfo {year} {2010})}\BibitemShut {NoStop}%
\bibitem [{\citenamefont {Anderson}(1958)}]{Anderson1958}%
  \BibitemOpen
  \bibfield  {author} {\bibinfo {author} {\bibfnamefont {P.~W.}\ \bibnamefont
  {Anderson}},\ }\bibfield  {title} {\enquote {\bibinfo {title} {{Absence of
  diffusion in certain random lattices}},}\ }\href {\doibase
  10.1103/PhysRev.109.1492} {\bibfield  {journal} {\bibinfo  {journal} {Phys.
  Rev.}\ }\textbf {\bibinfo {volume} {109}},\ \bibinfo {pages} {1492--1505}
  (\bibinfo {year} {1958})}\BibitemShut {NoStop}%
\bibitem [{\citenamefont {Billy}\ \emph {et~al.}(2008)\citenamefont {Billy},
  \citenamefont {Josse}, \citenamefont {Zuo}, \citenamefont {Bernard},
  \citenamefont {Hambrecht}, \citenamefont {Lugan}, \citenamefont {Cl\'ement},
  \citenamefont {Sanchez-Palencia}, \citenamefont {Bouyer},\ and\ \citenamefont
  {Aspect}}]{AspectNature2008}%
  \BibitemOpen
  \bibfield  {author} {\bibinfo {author} {\bibfnamefont {J.}~\bibnamefont
  {Billy}}, \bibinfo {author} {\bibfnamefont {V.}~\bibnamefont {Josse}},
  \bibinfo {author} {\bibfnamefont {Z.}~\bibnamefont {Zuo}}, \bibinfo {author}
  {\bibfnamefont {A.}~\bibnamefont {Bernard}}, \bibinfo {author} {\bibfnamefont
  {B.}~\bibnamefont {Hambrecht}}, \bibinfo {author} {\bibfnamefont
  {P.}~\bibnamefont {Lugan}}, \bibinfo {author} {\bibfnamefont
  {D.}~\bibnamefont {Cl\'ement}}, \bibinfo {author} {\bibfnamefont
  {L.}~\bibnamefont {Sanchez-Palencia}}, \bibinfo {author} {\bibfnamefont
  {P.}~\bibnamefont {Bouyer}}, \ and\ \bibinfo {author} {\bibfnamefont
  {A.}~\bibnamefont {Aspect}},\ }\bibfield  {title} {\enquote {\bibinfo {title}
  {{Direct observation of Anderson localization of matter waves in a controlled
  disorder}},}\ }\href {\doibase 10.1038/nature07000} {\bibfield  {journal}
  {\bibinfo  {journal} {Nature}\ }\textbf {\bibinfo {volume} {453}},\ \bibinfo
  {pages} {891--894} (\bibinfo {year} {2008})}\BibitemShut {NoStop}%
\bibitem [{\citenamefont {Balents}\ and\ \citenamefont
  {Fisher}(1997)}]{BalentsFisher1997}%
  \BibitemOpen
  \bibfield  {author} {\bibinfo {author} {\bibfnamefont {L.}~\bibnamefont
  {Balents}}\ and\ \bibinfo {author} {\bibfnamefont {M.~P.~A.}\ \bibnamefont
  {Fisher}},\ }\bibfield  {title} {\enquote {\bibinfo {title} {{Delocalization
  transition via supersymmetry in one dimension}},}\ }\href {\doibase
  10.1103/PhysRevB.56.12970} {\bibfield  {journal} {\bibinfo  {journal} {Phys.
  Rev. B}\ }\textbf {\bibinfo {volume} {56}},\ \bibinfo {pages} {12970--12991}
  (\bibinfo {year} {1997})}\BibitemShut {NoStop}%
\bibitem [{\citenamefont {Shelton}\ and\ \citenamefont
  {Tsvelik}(1998)}]{SheltonTsvelikPRB1998}%
  \BibitemOpen
  \bibfield  {author} {\bibinfo {author} {\bibfnamefont {D.~G.}\ \bibnamefont
  {Shelton}}\ and\ \bibinfo {author} {\bibfnamefont {A.~M.}\ \bibnamefont
  {Tsvelik}},\ }\bibfield  {title} {\enquote {\bibinfo {title} {{Effective
  theory for midgap states in doped spin-ladder and spin-Peierls systems:
  Liouville quantum mechanics}},}\ }\href {\doibase 10.1103/PhysRevB.57.14242}
  {\bibfield  {journal} {\bibinfo  {journal} {Phys. Rev. B}\ }\textbf {\bibinfo
  {volume} {57}},\ \bibinfo {pages} {14242--14246} (\bibinfo {year}
  {1998})}\BibitemShut {NoStop}%
\bibitem [{\citenamefont {Mkhitaryan}\ and\ \citenamefont
  {Raikh}(2011)}]{MkhitaryanPRL2011}%
  \BibitemOpen
  \bibfield  {author} {\bibinfo {author} {\bibfnamefont {V.~V.}\ \bibnamefont
  {Mkhitaryan}}\ and\ \bibinfo {author} {\bibfnamefont {M.~E.}\ \bibnamefont
  {Raikh}},\ }\bibfield  {title} {\enquote {\bibinfo {title} {{Localization
  properties of random-mass Dirac fermions from real-space renormalization
  group}},}\ }\href {\doibase 10.1103/PhysRevLett.106.256803} {\bibfield
  {journal} {\bibinfo  {journal} {Phys. Rev. Lett.}\ }\textbf {\bibinfo
  {volume} {106}},\ \bibinfo {pages} {256803} (\bibinfo {year}
  {2011})}\BibitemShut {NoStop}%
\bibitem [{\citenamefont {Sinai}(1982)}]{Sinai1982}%
  \BibitemOpen
  \bibfield  {author} {\bibinfo {author} {\bibfnamefont {Y.~G.}\ \bibnamefont
  {Sinai}},\ }\bibfield  {title} {\enquote {\bibinfo {title} {{The limiting
  behavior of a one-dimensional random walk in a random medium}},}\ }\href
  {\doibase 10.1137/1127028} {\bibfield  {journal} {\bibinfo  {journal} {Theory
  Probab. Appl.}\ }\textbf {\bibinfo {volume} {27}},\ \bibinfo {pages}
  {256--268} (\bibinfo {year} {1982})}\BibitemShut {NoStop}%
\bibitem [{\citenamefont {Bouchaud}\ \emph {et~al.}(1990)\citenamefont
  {Bouchaud}, \citenamefont {Comtet}, \citenamefont {Georges},\ and\
  \citenamefont {Le~Doussal}}]{Bouchaud1990}%
  \BibitemOpen
  \bibfield  {author} {\bibinfo {author} {\bibfnamefont {J.~P.}\ \bibnamefont
  {Bouchaud}}, \bibinfo {author} {\bibfnamefont {A.}~\bibnamefont {Comtet}},
  \bibinfo {author} {\bibfnamefont {A.}~\bibnamefont {Georges}}, \ and\
  \bibinfo {author} {\bibfnamefont {P.}~\bibnamefont {Le~Doussal}},\ }\bibfield
   {title} {\enquote {\bibinfo {title} {{Classical diffusion of a particle in a
  one-dimensional random force field}},}\ }\href {\doibase
  10.1016/0003-4916(90)90043-N} {\bibfield  {journal} {\bibinfo  {journal}
  {Ann. Phys.}\ }\textbf {\bibinfo {volume} {201}},\ \bibinfo {pages}
  {285--341} (\bibinfo {year} {1990})}\BibitemShut {NoStop}%
\bibitem [{\citenamefont {Comtet}\ and\ \citenamefont
  {Dean}(1998)}]{ComtetDean1998}%
  \BibitemOpen
  \bibfield  {author} {\bibinfo {author} {\bibfnamefont {A.}~\bibnamefont
  {Comtet}}\ and\ \bibinfo {author} {\bibfnamefont {D.~S.}\ \bibnamefont
  {Dean}},\ }\bibfield  {title} {\enquote {\bibinfo {title} {{Exact results on
  Sinai's diffusion}},}\ }\href {\doibase 10.1088/0305-4470/31/43/004}
  {\bibfield  {journal} {\bibinfo  {journal} {J. Phys. A}\ }\textbf {\bibinfo
  {volume} {31}},\ \bibinfo {pages} {8595} (\bibinfo {year}
  {1998})}\BibitemShut {NoStop}%
\bibitem [{\citenamefont {Yosprakob}\ and\ \citenamefont
  {Suwanna}(2016)}]{YosprakobSuwanna2016}%
  \BibitemOpen
  \bibfield  {author} {\bibinfo {author} {\bibfnamefont {A.}~\bibnamefont
  {Yosprakob}}\ and\ \bibinfo {author} {\bibfnamefont {S.}~\bibnamefont
  {Suwanna}},\ }\bibfield  {title} {\enquote {\bibinfo {title} {{Time evolution
  of Gaussian wave packets under Dirac equation with fluctuating mass and
  potential}},}\ }\href {https://arxiv.org/abs/1601.03827} {\bibfield
  {journal} {\bibinfo  {journal} {arXiv:1601.03827}\ } (\bibinfo {year}
  {2016})}\BibitemShut {NoStop}%
\bibitem [{\citenamefont {Steiner}\ \emph {et~al.}(1998)\citenamefont
  {Steiner}, \citenamefont {Fabrizio},\ and\ \citenamefont
  {Gogolin}}]{SteinerPRB1998}%
  \BibitemOpen
  \bibfield  {author} {\bibinfo {author} {\bibfnamefont {M.}~\bibnamefont
  {Steiner}}, \bibinfo {author} {\bibfnamefont {M.}~\bibnamefont {Fabrizio}}, \
  and\ \bibinfo {author} {\bibfnamefont {Alexander~O.}\ \bibnamefont
  {Gogolin}},\ }\bibfield  {title} {\enquote {\bibinfo {title} {{Random-mass
  Dirac fermions in doped spin-Peierls and spin-ladder systems: One-particle
  properties and boundary effects}},}\ }\href {\doibase
  10.1103/PhysRevB.57.8290} {\bibfield  {journal} {\bibinfo  {journal} {Phys.
  Rev. B}\ }\textbf {\bibinfo {volume} {57}},\ \bibinfo {pages} {8290--8306}
  (\bibinfo {year} {1998})}\BibitemShut {NoStop}%
\bibitem [{\citenamefont {Succi}\ and\ \citenamefont
  {Benzi}(1993)}]{SucciBenzi1993}%
  \BibitemOpen
  \bibfield  {author} {\bibinfo {author} {\bibfnamefont {S.}~\bibnamefont
  {Succi}}\ and\ \bibinfo {author} {\bibfnamefont {R.}~\bibnamefont {Benzi}},\
  }\bibfield  {title} {\enquote {\bibinfo {title} {{Lattice Boltzmann equation
  for quantum mechanics}},}\ }\href {\doibase 10.1016/0167-2789(93)90096-J}
  {\bibfield  {journal} {\bibinfo  {journal} {Physica D}\ }\textbf {\bibinfo
  {volume} {69}},\ \bibinfo {pages} {327--332} (\bibinfo {year}
  {1993})}\BibitemShut {NoStop}%
\bibitem [{\citenamefont {Palpacelli}\ and\ \citenamefont
  {Succi}(2008)}]{PalpacelliSucciPRE2008}%
  \BibitemOpen
  \bibfield  {author} {\bibinfo {author} {\bibfnamefont {S.}~\bibnamefont
  {Palpacelli}}\ and\ \bibinfo {author} {\bibfnamefont {S.}~\bibnamefont
  {Succi}},\ }\bibfield  {title} {\enquote {\bibinfo {title} {{Quantum lattice
  Boltzmann simulation of expanding Bose-Einstein condensates in random
  potentials}},}\ }\href {\doibase 10.1103/PhysRevE.77.066708} {\bibfield
  {journal} {\bibinfo  {journal} {Phys. Rev. E}\ }\textbf {\bibinfo {volume}
  {77}},\ \bibinfo {pages} {066708} (\bibinfo {year} {2008})}\BibitemShut
  {NoStop}%
\bibitem [{\citenamefont {Fillion-Gourdeau}\ \emph {et~al.}(2013)\citenamefont
  {Fillion-Gourdeau}, \citenamefont {Herrmann}, \citenamefont {Mendoza},
  \citenamefont {Palpacelli},\ and\ \citenamefont
  {Succi}}]{FillionGourdeauPRL2013}%
  \BibitemOpen
  \bibfield  {author} {\bibinfo {author} {\bibfnamefont {F.}~\bibnamefont
  {Fillion-Gourdeau}}, \bibinfo {author} {\bibfnamefont {H.~J.}\ \bibnamefont
  {Herrmann}}, \bibinfo {author} {\bibfnamefont {M.}~\bibnamefont {Mendoza}},
  \bibinfo {author} {\bibfnamefont {S.}~\bibnamefont {Palpacelli}}, \ and\
  \bibinfo {author} {\bibfnamefont {S.}~\bibnamefont {Succi}},\ }\bibfield
  {title} {\enquote {\bibinfo {title} {{Formal analogy between the Dirac
  equation in its Majorana form and the discrete-velocity version of the
  Boltzmann kinetic equation}},}\ }\href {\doibase
  10.1103/PhysRevLett.111.160602} {\bibfield  {journal} {\bibinfo  {journal}
  {Phys. Rev. Lett.}\ }\textbf {\bibinfo {volume} {111}},\ \bibinfo {pages}
  {160602} (\bibinfo {year} {2013})}\BibitemShut {NoStop}%
\bibitem [{\citenamefont {Dellar}\ \emph {et~al.}(2011)\citenamefont {Dellar},
  \citenamefont {Lapitski}, \citenamefont {Palpacelli},\ and\ \citenamefont
  {Succi}}]{DellarPRE2011}%
  \BibitemOpen
  \bibfield  {author} {\bibinfo {author} {\bibfnamefont {P.~J.}\ \bibnamefont
  {Dellar}}, \bibinfo {author} {\bibfnamefont {D.}~\bibnamefont {Lapitski}},
  \bibinfo {author} {\bibfnamefont {S.}~\bibnamefont {Palpacelli}}, \ and\
  \bibinfo {author} {\bibfnamefont {S.}~\bibnamefont {Succi}},\ }\bibfield
  {title} {\enquote {\bibinfo {title} {{Isotropy of three-dimensional quantum
  lattice Boltzmann schemes}},}\ }\href {\doibase 10.1103/PhysRevE.83.046706}
  {\bibfield  {journal} {\bibinfo  {journal} {Phys. Rev. E}\ }\textbf {\bibinfo
  {volume} {83}},\ \bibinfo {pages} {046706} (\bibinfo {year}
  {2011})}\BibitemShut {NoStop}%
\bibitem [{\citenamefont {Succi}(2015)}]{SucciEPL2015}%
  \BibitemOpen
  \bibfield  {author} {\bibinfo {author} {\bibfnamefont {S.}~\bibnamefont
  {Succi}},\ }\bibfield  {title} {\enquote {\bibinfo {title} {{Lattice
  Boltzmann 2038}},}\ }\href {\doibase 10.1209/0295-5075/109/50001} {\bibfield
  {journal} {\bibinfo  {journal} {EPL}\ }\textbf {\bibinfo {volume} {109}},\
  \bibinfo {pages} {50001} (\bibinfo {year} {2015})}\BibitemShut {NoStop}%
\end{thebibliography}

%

\end{document}